\documentclass[manuscript]{aastex}

\usepackage{graphicx}
\usepackage{amssymb}

\usepackage[dvipsnames]{color}

\shorttitle{Carbon ionization states and the cosmic far-UV background with HeII absorption}
\shortauthors{Vasiliev, Sethi, Nath}

\begin{document}


\newcommand{\3}{\ss}
\newcommand{\n}{\noindent}
\newcommand{\eps}{\varepsilon}
\def\be{\begin{equation}}
\def\ee{\end{equation}}
\def\ba{\begin{eqnarray}}
\def\ea{\end{eqnarray}}
\def\de{\partial}
\def\msun{M_\odot}
\def\div{\nabla\cdot}
\def\grad{\nabla}
\def\rot{\nabla\times}
\def\ltsima{$\; \buildrel < \over \sim \;$}
\def\simlt{\lower.5ex\hbox{\ltsima}}
\def\gtsima{$\; \buildrel > \over \sim \;$}
\def\simgt{\lower.5ex\hbox{\gtsima}}
\def\etal{{et al.\ }}
\def\red{\textcolor{red}} 
\def\blue{\textcolor{blue}}


\title{Carbon ionization states and the cosmic far-UV background with HeII absorption}

\author{Evgenii O. Vasiliev}
\affil{Institute of Physics, Southern Federal University, Stachki Ave. 194, Rostov-on-Don, 344090, Russia}
\email{eugstar@mail.ru}

\author{Shiv K. Sethi}
\affil{Raman Research Institute, Sadashiva Nagar, Bangalore 560080, India}
\email{sethi@rri.res.in}

\author{Biman B. Nath}
\affil{Raman Research Institute, Sadashiva Nagar, Bangalore 560080, India}
\email{biman@rri.res.in}

\begin{abstract}
We  constrain
the spectrum of the cosmic ultraviolet background radiation 
by fitting
the observed abundance ratios carbon ions at $z\sim 2\hbox{--}3$
with those expected from different models of the background radiation. 
We use the recently calculated modulation of the background radiation 
between 3 and 4 Ryd due to resonant line absorption by intergalactic HeII,
and determine the ratios of CIII to CIV expected at these redshifts, as
functions of metallicity, gas density and temperature. Our analysis of
the observed ratios shows that 'delayed reionization' models, which
assume a large fraction of HeII
at $z\sim3$, is not favored by data. Our results suggest that HeII
reionization was inhomogeneous, consistent with the predictions from
recent simulations.
\end{abstract}


\section{Introduction}
Cosmological simulations of the intergalactic medium (IGM) have been used
in the recent years to determine many important aspects of structure formation 
and reionization. 
One important input for these simulations, the cosmic ultraviolet 
background radiation, however, has remained uncertain. Our knowledge
of the cosmic UV background, its intensity and spectrum and evolution
with time, still lacks the precision that is needed to make the 
interpretations from cosmological simulations robust.

It is believed that HeII reionization occurred much after the reionization
of HI and HeI, with the help of hard UV photons from quasars, whose activity
peaked around redshift $z\sim 3$. This is because of the relatively high
ionization threshold of HeII (54.4 eV) \citep[e.g.,][]{madau94,sokasian}. 
Some recent observations may have found evidences for HeII
Gunn-Peterson trough in the spectra of quasars at $z \ge 2.8$
\citep[e.g.,][]{jakobsen,schaye,smette}.
Numerical simulations of HeII reionization \citep{mcquinns} also
predict that the
process is likely to be inhomogeneous, and accompanied by heating of the IGM. 
There are some evidences for such a heating from recent observations, although
the interpretations are not yet clear \citep[e.g.,][]{ricotti,schaye,theuns,bernardi}.

Recently, \citet{mh09} pointed out an important aspect of the
evolution of UV background radiation in the context of HeII reionization.
They calculated the effect of resonant absorption by HeII Lyman series
that is likely to significantly 
change the shape of the UV background radiation
between 3 and 4 Ryd, whose magnitude depends
on the abundance of HeII in the IGM. This modulation can be an important
probe of the UV background radiation, when combined with the
observations of 
absorption lines from metal ions such as CIII, whose ionization potential
lies between energies corresponding to 3 and 4 Ryd.

In this paper, we use the data from \citet{agafonova} of column
densities of CIII and CIV in the redshift range of 
$z \sim 2\hbox{--}3$, to constrain the spectrum of the UV background radiation.
\citet{agafonova} used their observations to recover the shape of the
spectrum with a Monte Carlo procedure, but their assumption for the shape of 
the spectrum did not include the possible attenuation between 3 and 4 Ryd. 
They claimed a sharp reduction in the flux of the UV background
between 3 and 4 Ryd, which was interpreted as a sign of HeII Gunn-Peterson
(GP) effect. In this regard, it is important to study the ionic ratios, 
especially those involving CIII, whose ionization potential falls between 3 and 4 Ryd,
and the effect of background radiations of different spectrum on these ratios.
One of the predictions made by \citet{mh09} was that the
modulation of the spectrum by HeII absorption would significantly
change the abundance ratio of CIII/CIV at $z\sim 3$, and a large CIII/CIV
ratio could be explained without any need to invoke the HeII GP effect. 
We test these ideas in this paper.

\section{Calculations of ionic states}
We study the thermal and ionization evolution of gas exposed to external ionizing 
radiation. Our calculation of the evolution of gas is similar to \citet{gs07,gs09}. 
A detailed description of thermal and ionization evolution of gas 
can be found in \citet{v10}. Here we briefly describe the method of 
calculation, present several tests of our code and discuss the choice of 
initial conditions.

\subsection{The description of the code}

A gas parcel is assumed to be optically thin for the ionizing radiation. 
We consider the time-dependent equations for all ionization states of H, He, C, N, 
O, Ne and Fe, including all relevant atomic processes.
Namely, we take into account the following major processes: photoionization, collisional 
ionization, radiative and dielectronic recombination as well as charge transfer in 
collisions with hydrogen and helium atoms and ions. Atomic data for the photoionization 
cross section are adopted from \citet{verner96,vya95}, Auger 
effect probabilities taken from \citet{kaastra}, the recombination rate, for 
radiative recombination taken from \citet{vf96}, \citet{pequignot}, 
\citet{aray} and \citet{shull82}, 
for dielectronic recombination adopted from \citet{mazzotta}, and for the collisional 
ionization rate adopted from \citet{voronov}, for the charge transfer rates of ionization and 
recombination with hydrogen and helium adopted from \citet{arot} and \citet{kingdon}.

This system of time-dependent ionization state equations should be complemented by the 
temperature equation, which accounts for all relevant cooling and heating processes. 
Here we assume that a gas parcel cools isochorically, i.e. the cooling time is
shorter than any dynamical time scale.
Note that in the temperature equation we neglect the change in the number of particles in the 
system (which for a fully ionized plasma of hydrogen and helium remains approximately constant). 

The total cooling and heating rates are calculated using the photoionization code 
CLOUDY \citep[ver. 08.00,][]{cloudy}. More specifically, we incorporate into the CLOUDY 
code a given set of all ionic fractions $X_i$ calculated at temperature $T$, gas 
number density $n$ and external ionization flux $J(\nu)$ and obtain the corresponding 
cooling and heating rates. 
For the solar metallicity case we adopt the abundances reported by \citet{asplund},
except Ne for which the enhanced abundance is adopted \citep{drake}.
In all our calculations we assume the helium mass fraction to be $Y_{\rm He} = 0.24$.

We solve a set of 68 coupled equations (67 for all ionization states of
the elements H, He, C, N, O, Ne and Fe and one for temperature) 
using a Variable-coefficient Ordinary Differential Equation solver \citep{dvode}. 
The time step is chosen to be the minimum value between ($0.1 t_{ion}, 0.01 t_{cool}$),
where $t_{ion}$ and $t_{cool}$ are the ionization and cooling times correspondingly.

\begin{figure}
\vspace{24pt}
\includegraphics[width=80mm]{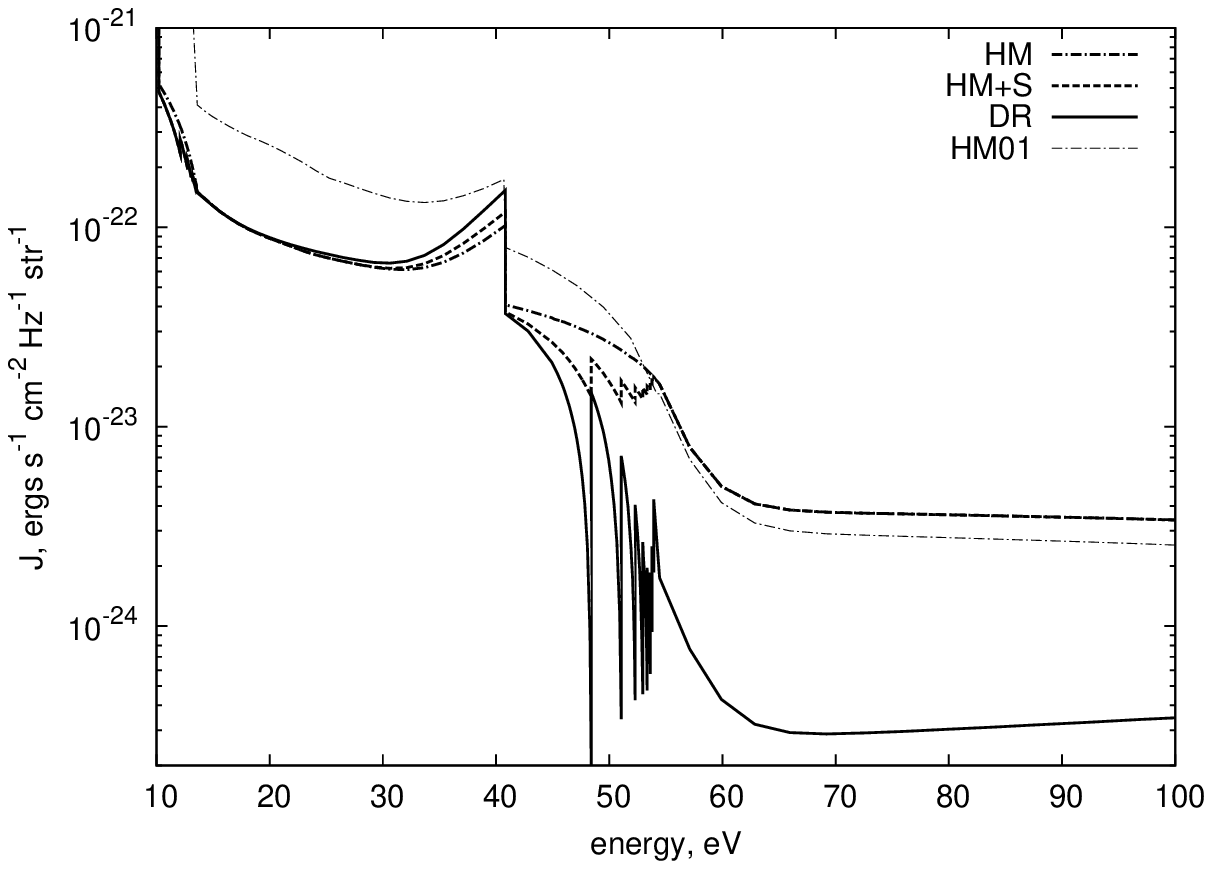}  
\caption{
The far ultraviolet part of the extragalactic ionizing spectra at {  $z=3$ } \citep{mh09}:
the spectrum without absorption in the HeII resonant lines -- $HM$ model -- thick dot-dashed line, 
the spectrum with the sawtooth modulation -- $HM+S$ model (HeII/HI=35 in optically thin absorbers) --
thick dashed line, and the spectrum model where the ratio HeII/HI were artificially increased to 250 
-- $DR$ (delayed reionization) model -- thick solid line. 
For comparison the \citet{hm96,hm01} spectrum ($HM01$) for quasar-only (thin dot-dashed line) 
is added.
         }
\label{figspec}
\end{figure}

We study the ionization and thermal evolution of a lagrangian element of cooling gas exposed to 
extragalactic spectra calculated by using the radiative transfer in the cosmological
simulations \citep{mh09} with and without the absorption in the He II resonant lines. 
Figure~\ref{figspec} shows the 
extragalactic spectra used here. A brief description of the spectra is presented in Section 3.
In our simulations we take into account the radiation spectrum from 1 to $10^4$ eV.

The evolution of the gas certainly depends on a choice of the initial condition of gas, namely, 
the initial ionic composition and temperature. 
We consider such a dependence below (section 2.3). But first,
we present several tests of the method of calculation.
In these tests we start our calculations from temperature $T = 10^8$~K with the collisional 
equilibrium ion abundances. 
During the calculation the gas number density is assumed to remain 
unchanged due to any dynamical 
or chemical processes.
We stop our calculations when photoionization heating rate differs from cooling rate by less than 
5\% ($|\Gamma-\Lambda| /\Lambda < 0.05$). The 
ionic composition is almost "frozen" at this moment. Note that with
 such a stopping criterion 
a gas still remains "time-dependent" at the end of the computation, as it has not yet reached 
thermal and ionization equilibrium.

\subsection{Tests of the code}

\begin{figure}
\vspace{24pt}
\includegraphics[width=80mm]{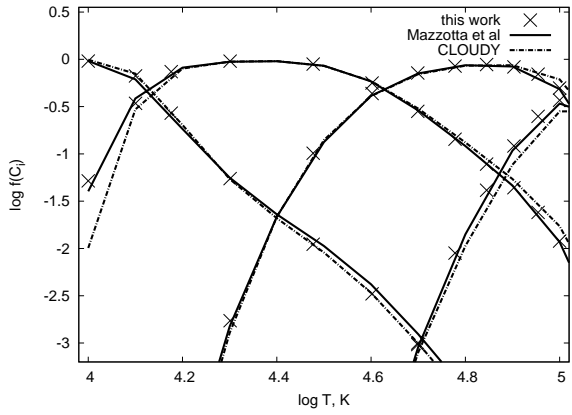}  
\caption{
         The collisional equilibrium carbon I-IV state fractions 
         for our code (crosses) are superposed on the
         original results by \citet{mazzotta} (solid lines) and the 
         standard collisional equilibrium test of CLOUDY (dash-dotted lines).
         }
\label{figcie}
\end{figure}

\begin{figure}
\vspace{24pt}
\includegraphics[width=80mm]{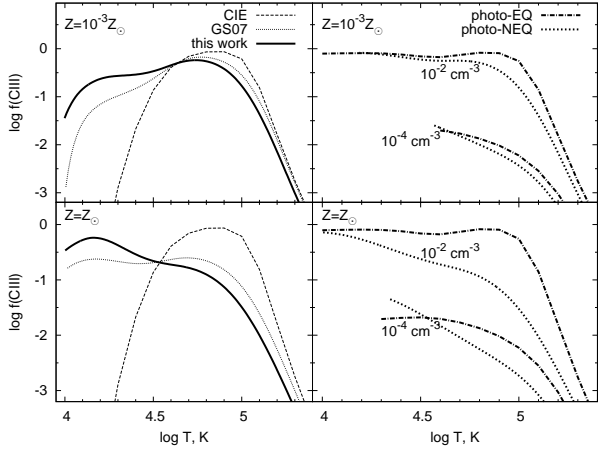}  
\caption{
         { 
         The CIII ion fraction for $10^{-3}Z_\odot$ (upper panels) and solar (lower panels)
         metallicities for collisional (left panels) and photoionization (right panels) cases.
         The collisional time-dependent ion fraction obtained using our code is shown by thick 
         solid line and that taken from \citet{gs07} is depicted by thin solid
         line (GS07). 
         In both left panels the CIE fraction of CIII obtained from CLOUDY is shown by 
         thin dashed line.
         The time-dependent photoionization (thick dot lines, "photo-NEQ") and photo
         equilibrium (thick dash-dot lines, "photo-EQ") fractions are shown for $n = 10^{-4}$~cm$^{-3}$ 
         by lower thick lines and for $n = 10^{-2}$~cm$^{-3}$ by upper thick lines, correspondingly. 
         In the photoionization calculations the extragalactic ionizing spectrum without
         any modulation ($HM$ model in Figure~\ref{figspec}) at $z=3$ \citep{mh09}
         is assumed.
         }
         }
\label{figcompare}
\end{figure}

Figures~\ref{figcie}-\ref{figcompare} present the main tests of our code used in the calculations 
at the temperature range interested in this paper. Figure~\ref{figcie} shows the carbon ionization
fractions I-IV in the collisional equilibrium obtained from our code (crosses of various types) and 
original results by \citet{mazzotta} as well as the standard CIE test of the CLOUDY code. One 
can see a good coincidence between the results of our code and those obtained in the previous works. 

We show the comparison between the cases of photoionization equilibrium and time-dependent
photoionization as well as collisional time-dependent and equilibrium
with regard to CIII fraction in Figure~\ref{figcompare}. The case of photoionization
equilibrium is shown with thick dash-dot lines and that of the 
time-dependent photoionization is shown by thick dot lines. We use
a density of $n = 10^{-4}$cm$^{-3}$ (lower thick lines) and 
$n=10^{-2}$cm$^{-3}$ (upper thick lines). The upper panels show
the results for $10^{-3}Z_\odot$ and those for the solar metallicity are
shown in the lower panels. In the photoionization calculations the extragalactic 
ionizing spectrum without any modulation -- $HM$ model, (see Figure~\ref{figspec}) 
at $z=3$ \citep{mh09} is assumed.

Figure~\ref{figcompare} also shows the collisional time-dependent CIII fraction obtained using 
our code (thick solid line) and that taken from \citet{gs07} (thin solid line), as 
well as the CIE fractions obtained from CLOUDY (thin dash line). Firstly, one notes the difference 
of our results from the \citet{gs07} data. 
This difference can be explained by the fact that we use a different set of atomic data (radiative 
and dielectronic recombination rates) for several ions. We use the data for radiative recombination 
from Verner \& Ferland (1996), which partially included the refitted data from previous works (see 
above for full list of papers), and for dielectronic recombination from \citet{mazzotta}. 
\citet{gs07} used more recent data of Badnell and coauthors \citep[see references in][]{gs07,bryans}.
However, our CIE results are close
to those obtained in CLOUDY code, which traditionally includes the most recent atomic
data\footnote{http://wiki.nublado.org/wiki/RevisionHistory}. So we assume that such difference
may be mainly explained by small changes in thermal evolution. For example, it is well known that 
the dielectronic recombination depends exponentially on temperature, so we expect that for higher
temperature and lower metallicity this difference becomes smaller (see Figure~\ref{figcompare}).

Secondly, we should note that both equilibrium and time-dependent collisional ionic 
composition has no dependence on gas density, because all processes are two-body.
In contrast, the ionic composition in the photoionization case strongly depends on 
gas density. Moreover, the ionizing radiation forces the ionic composition of gas to 
settle on to equilibrium. In low density gas the ionic composition in the time-dependent 
photoionization case is expected to differ strongly from that in collisional both equilibrium 
and time-dependent cases, but it tends to the ionic fractions in the photoequilibrium. 
In high density gas we expect that the time-dependent photoionization ionic composition tends 
to the time-dependent collisional one, whereas the photoequilibrium ionic fractions should 
be close to those in the CIE. 
In addition, the ionic composition strongly depends on 
metallicity. For low metallicity the difference between time-dependent and equilibrium ionic 
fractions is expected to be small, but it increases with metallicity. 

In Figure~\ref{figcompare}
one can see that the CIII fraction in the time-dependent photoionization model does not fully 
coincide in the whole temperature range with either the photoionization equilibrium or pure 
collisional models. For example, at low metallicity, $Z = 10^{-3}~Z_\odot$, for high density 
value, $n = 10^{-2}$~cm$^{-3}$, the time-dependent photoionization CIII fraction shows some 
difference from the photo equilibrium case. In contrast, for $n = 10^{-4}$~cm$^{-3}$, 
the CIII fractions in the time-dependent photoionization and photo-equilibrium cases are 
close to each other in the whole temperature range. For solar metallicity, the difference 
between time-dependent and equilibrium photoionization becomes more significant for both number 
density values.
Also one should note from the right panels of Figure~\ref{figcompare} there is almost no 
dependence on metallicity in equilibrium photoionization case. The difference between 
the CIII fraction for $10^{-3}~Z_\odot$ and solar metallicities is only in the value of
temperature reached in the cooling process: for higher cooling rate (solar metallicity)
this temperature is lower.

As it is expected for higher density value, $n = 10^{-2}$cm$^{-3}$,
the CIII photoionization fractions are close to the collisional ones 
in both equilibrium and time-dependent cases in a wide 
temperature range. For example, the ionic fractions in 
time-dependent collisional (thick solid line) and time-dependent
photoionization cases (thick dotted line) almost coincide for 
${\rm log} T \simgt 4.7$ at $Z = 10^{-3}Z_\odot$ and for 
${\rm log} T \simgt 4.18$ at solar metallicity. 
In summary, both time-dependent collisional and photoequilibrium in the presence of a 
significant ionizing radiation flux like the extragalactic background models
produce the ionic composition, which differs from that in time-dependent photoionization model. 
Due to the very non-linear dependence of ionic fractions on temperature, metallicity and 
density it is difficult to say where it is possible to use time-dependent collisional or 
photoequilibrium models, so in the present work we use the more complex, but more adequate 
time-dependent photoionization model.

\subsection{Dependence on the initial temperature}

Here we should consider a choice of the initial temperature and ionic composition for our study
of absorbers at the redshift range $z=2-3$. 
Many cosmological simulations \citep{dave01,cen,bertone}
show that the intergalactic gas at $z=2-3$ mainly resides in the diffuse phase with a typical 
temperature $T\simlt 10^5$K and overdensity $\delta<1000$, and a small part of gas (in the 
best model this fraction is less than 5\%, see e.g. Dav\'e et al. 2001) can be found in the 
warm-hot phase with 
$10^5$K$\simlt T\simlt 10^7$K. It is expected that a gas enriched by metals and expelled from 
galaxies would have passed through strong shock waves. Since the typical velocity of galactic winds
varies from several tens to hundreds km~s$^{-1}$, the intergalactic gas enriched by metals is 
likely to go through such shock waves. We assume that such a gas was initially heated by shocks
up to the temperature $T\gg 10^4$K. We are interested in subsequent evolution of this shock-heated gas.
Here also we should note, firstly, that the heating rate produced by the strong ionizing background 
at $z=2-3$ does not allow low-density gas with $n\sim 10^{-4}$cm$^{-3}$ to cool effectively below 
$T\sim 3.5\times 10^4$K, and secondly, that here we do not include adiabatic cooling due to the 
expansion of the universe, which dominates over the radiative cooling for very low density gas, 
$n\sim 10^{-5}$~cm$^{-3}$ at $z=3$. 

In the following set of calculations we study the dependence of the ionic composition evolution 
of a gas on the initial temperature $T_i$. We vary the initial temperature in a wide range 
$T_i = 5\times 10^4 - 10^8$K. Here we follow the evolution of gas irrespective of
the time needed to cool from $T_i$ to the temperature value, when our stopping criterion
(mentioned above) is reached. Further we compare the cooling time with the comoving Hubble time
at $z=2-3$. The gas cooling is strongly determined by both initial conditions of 
the gas (density, metallicity and temperature) and the UV background radiation. 
We assume the initial ionic composition corresponds to $T_0 = 2\times 10^4$K. Taking such 
conditions we simulate a gas parcel with $T_0 = 2\times 10^4$K that initially passes through 
a strong shock wave front with $T_s\gg T_0$. Because of the short ionization timescale of
metals (ionization fraction of hydrogen reaches almost unity for $T_0 = 2\times 10^4$K in both 
CIE and time-dependent cases) the ionic composition of a gas passed through a shock front with 
temperature $T_s\gg T_0$ 
is expected to tend to that with $T_i = 10^8$K at $T\simlt T_s$. 
Certainly, if the initial temperature is low enough, then the ionic composition should strongly 
depend on the initial temperature value. 

\begin{figure}
\vspace{24pt}
\includegraphics[width=80mm]{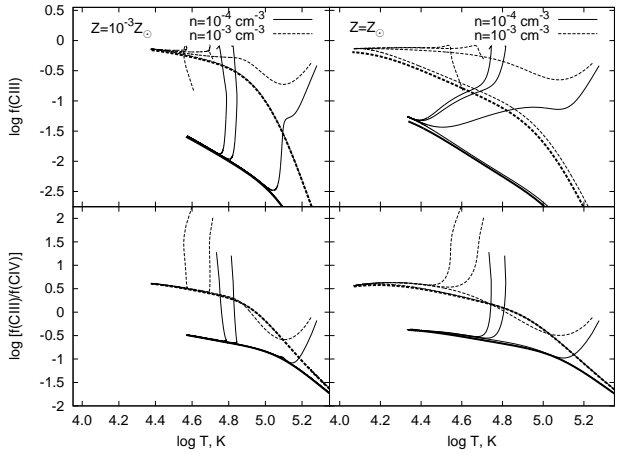}  
\caption{
         {\it Upper panels.} The CIII ion fraction for gas with $Z=10^{-3}Z_\odot$ (left panel) 
         and with solar metallicity (right panel), and 
         $n = 10^{-4}$~cm$^{-3}$ (solid lines), $n = 10^{-3}$~cm$^{-3}$ (dashed lines)
         for different initial temperature values. Thick lines correspond to
         the initial temperature $T = 10^8$K, thin lines present the evolutionary tracks 
         for initial temperature $T = 5\times 10^4, 10^5, 4\times 10^5, 10^6$K 
         (from top/left to bottom/right).
         {\it Lower panels.} The CIII/CIV ratio for gas with the same parameters as
         in the {\it upper} panels.
         }
\label{figtini}
\end{figure}

Figure~\ref{figtini} presents the CIII fraction ({\it upper} panels) and the CIII/CIV ratio 
({\it lower} panels) for different starting temperature values and the ionization composition 
initially corresponding to that in the CIE at $T_0 = 2 \times 10^4$K. 
In the upper panels (the left is for $Z=10^{-3}Z_\odot$, the right is for the solar metallicity)
the significant deviations from the CIII fraction for the evolutionary track with $T_i = 10^8$K 
is found for $T< 10^6$K, whereas the CIII fraction for tracks with $T_i = 10^6$K and $T_i = 10^8$K 
almost coincides. The time needed for 
cooling a gas from $T_i\sim 10^6$K with $n\sim 10^{-4}$~cm$^{-3}$ and $Z\sim 10^{-3}Z_\odot$ 
is comparable with the comoving Hubble time at $z=2$ ($t_H(z=2) \simeq 1.1\times 10^{17}$~s). 
The cooling times from $T_i\sim 10^5$K is slightly shorter than the Hubble time at $z=3$,
but the CIII fraction strongly differs from that for $T_i = 10^6$K.
But here we are interested in the ionic ratios, and not in column densities of individual ionic species. 

In the {\it lower} panels of Figure~\ref{figtini} the CIII/CIV ratio demonstrates a weak dependence on 
the initial temperature. The ratio for $T_i = 4\times 10^5$K almost coincides with that for 
higher temperature values. For the lowest temperature value, $T_i = 5\times 10^4$K, considered 
in this set of calculations, the ionic ratio settles to the common trend almost at the 
same temperature $T\simlt 5\times 10^4$K for $n=10^{-4}$~cm$^{-3}$ (in this case at the beginning
the photoheating is significant) and at $T\simlt 3.2\times 10^4$K for $n=10^{-3}$~cm$^{-3}$. One can
see the vertical parts of the CIII/CIV ratio tracks. However, the time needed for settling on the 
common trend (the time at the vertical part of the track) is significantly lower than the time
needed for reaching the stopping criterion. For instance, for a gas with $n=10^{-4}$~cm$^{-3}$
and $Z=10^{-3}Z_\odot$ such a timescale is about $2\times 10^{15}$s in case of the initial temperature
$T_i = 5\times 10^4$K, whereas the time of the calculation before the stopping criterion
reached is more than order greater, $\sim 5.2\times 10^{16}$s$\simeq 0.7t_H(z=3)$. The timescales
for higher density or metallicity are smaller. Thus, for $T\simlt 4\times 10^4$K 
\citep[that is consistent with the line widths inferred by][]{agafonova} 
the CIII/CIV ratio in a gas with initial temperature 
$T_i\simgt 5\times 10^4$K almost coincides with that for $T_i = 10^6 - 10^8$K, and the timescale
of calculation needed for reaching our stopping criterion is smaller than the comoving Hubble time
at $z=2-3$. In light of this, we can start our calculations from temperature 
$T = 10^6$~K
and consider the obtained CIII/CIV ratio. 
We should emphasize that we are interested in the temperature range $T\simlt 4\times 10^4$K, 
where the ionic ratios for $T_i = 5\times10^4-10^8$K are very close (see lower panels of
Figure~\ref{figtini}), thus, our conclusions are almost independent on the initial temperature value, 
if this value is greater than $5\times 10^4$K.
Further analysis of the dependence of ionic composition on the initial conditions is out of scope 
of this paper and will be done elsewhere.

\section{CIII/CIV ratio}
In this section we study the influence of the extragalactic ionizing background with
and without HeII absorption on CIII/CIV ratio. Nonequilibrium ionization states are 
calculated for ultraviolet (UV) spectra at redshifts $z = 1.87, 2.48, 2.65, 2.83, 3$
(all spectral data were kindly provided by F. Haardt) for three models presented in 
\citet{mh09}, hereafter MH09 (the abbreviations of models are the same as in MH09): 
a) absorption in the HeII resonant lines was neglected -- $HM$ model, b) the sawtooth 
modulation was added -- $HM+S$ model, where HeII/HI=35 in optically thin absorbers, 
and c) the model where the ratio HeII/HI were artificially increased to 250 -- $DR$
(delayed reionization) model (for details see MH09). 

Note that the MH09 spectra differ 
from the \citet{hm96,hm01} spectra. First, MH09 spectra include quasar 
contribution only, and, second, in the UV background calculations a QSO luminosity 
function from \citet{hopkins} is used, which produces a very steep decline of the 
ionizing flux at high redshifts.
Figure~\ref{figspec} shows the far UV part of the MH09 spectra for $z = 2.48$. 

\begin{figure}
\vspace{24pt}
\includegraphics[width=80mm]{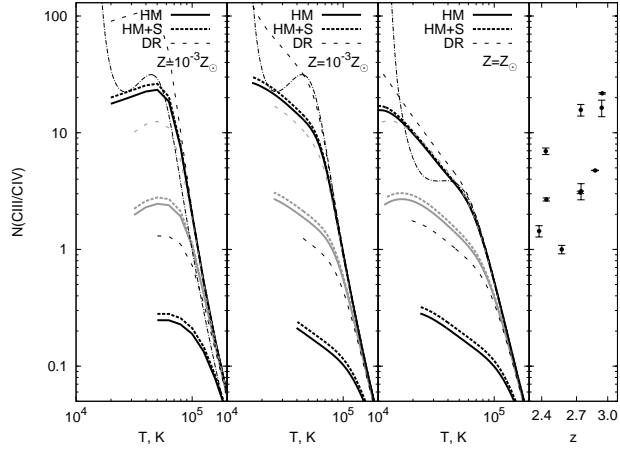}
\caption{
{\it Left panel.} The equilibrium CIII/CIV ratio for the \citet{mh09}
spectrum at $z=2.48$ for $Z=10^{-3}Z_\odot$ metallicity 
is plotted as a function of gas temperature for various cases. 
The solid lines 
depict the ratio for the "HM" extragalactic spectrum, dashed lines -- "HM+S", 
and dotted lines -- "DR"  models for three values of number density 
$n=10^{-4},\ 10^{-3},\ 10^{-2}$~cm$^{-3}$ (from bottom to top). 
{ For clarity, the models with $n=10^{-3}$~cm$^{-3}$ are shown by gray lines.}
Thin dot-dash line corresponds to the ratio in the collisional limit: 
for $Z=10^{-3}Z_\odot$.
{\it Left middle panel.} The same as in the {\it left} panel, but for nonequilibrum models.
{\it Right middle panel.} 
The same as in the {\it left middle} panel, but for solar metallicity.
{\it Right panel.} The observational data of CIII and CIV column densities for 
10 absorbers in the redshift range $2 \le z \le 3$ as given by \citet{agafonova}.
               }
\label{figc}
\end{figure}

The left panel of Figure~\ref{figc} shows CIII/CIV ratios in the equilibrium calculation 
(using CLOUDY) for $Z=10^{-3}Z_\odot$ (but there is no dependence on metallicity in the
equilibrium, see Section 2.2 and right panels of Figure~\ref{figcompare}). 
The two middle panels of Figure~\ref{figc} shows CIII/CIV ratios 
in the nonequilibrium calculation (using our program for nonequilibrium calculation)
for $Z=10^{-3}Z_\odot$ 
and solar metallicities, correspodingly, for UV background spectrum at $z=2.48$. The difference 
between models without ($HM$) and with ($HM+S$) the inclusion of the sawtooth modulation is small
for the whole range of metallicities and densities. The ratios for the $HM$ and $HM+S$ 
models differ from each other only by a factor of $\sim 1.3$ for low density at $T\simlt 10^5$~K and they
almost concide for $n=10^{-2}$cm$^{-3}$ or at $T\simgt 10^5$~K. In the $DR$ model the ratio 
$N({\rm CIII})/N({\rm CIV})$ is greater than those in the $HM$ and $HM+S$ models by a factor of $\sim10$
at $T\simlt 10^5$~K. This factor decreases with the increase of density and metallicity. For 
example, for $n=10^{-2}$cm$^{-3}$ a significant difference can be found only at $T\simlt 
6\times 10^4$~K for $Z=10^{-3}Z_\odot$ and $T\simlt 2\times 10^4$~K for solar metallicity.

For high density ($n\simgt 10^{-2}$~cm$^{-3}$), the ratio is close to that in the collisional 
limit at $T\simgt 5\times 10^4$~K. One can therefore conclude that large sawtooth modulation as 
present in the $DR$ model leads to higher abundances of CIII ions by a factor of 10--1.3 for gas 
density $n = 10^{-4} -10^{-2}$~cm$^{-3}$, respectively. The increase of CIII and CIV abundances in the
$DR$ model arises from the decrease of CV and CVI abundances, because of lower ionization flux in the
energy range $\sim 50-500$~eV in comparison with two other models, so the growth of the 
CIII abundance is higher than that of CIV making CIII/CIV ratio higher. 

Our results firstly show that the ratio of CIII/CIV is not significantly changed by 
the sawtooth modulation of HeII, as supposed by \citet{mh09}. Since the 
ionization threshold of CIII ($47.9$ eV) lies within the Ly$\beta$ absorption feature 
from HeII, and that of CIV lies well beyond the range where HeII resonant absorption 
changes the spectrum, it was expected that observations of the CIII/CIV ratio along 
with the theoretically calculated spectrum would be a good probe of the physical
conditions in the IGM at $z\sim 3$. Our detailed calculations confirm the \citet{mh09}
claim that the ionization rate of CIII does not significantly differ between the "HM" and 
"HM+S" spectra, although the "DR" spectrum does make a difference. In other
words, the CIII/CIV ratio is a good probe only for distinguishing between
the extreme cases of "HM" or "DR", and not for probing the attenuation
caused by standard abundances of HeII in the IGM.

\section{Statistical analysis}
\citet{agafonova} give measurements of CIII and CIV column densities for 
10 absorbers in the redshift range $2 \le z \le 3$. We use the ratio 
$R = N({\rm CIII})/N({\rm CIV})$ to constrain the ionizing radiation at far-UV range
(Table~1).

For our analysis we consider three  models of ionizing radiation, as discussed in the 
previous section, in the redshift range $2< z <3$ , the  metallicity in the range 
$Z= 10^{-3}\hbox{--}1$, the number density in the range $n = 10^{-4}\hbox{--}10^{-2}$~cm$^{-3}$, 
and temperature $T \le 4 \times 10^4$~K. This range of temperature is  consistent with the line 
widths inferred by \citet{agafonova}.

The measured ratio $R_{\rm ob}$ depends on the parameters $n, T,Z$ and the model of 
ionization. One expects the $Z$, $n$ and $T$ to vary appreciably from one absorber  
to another and therefore the usual $\chi^2$ approach cannot be applied. Only the 
background ionizing flux can be  assumed to vary slowly with redshift. In the light 
of these assumptions, we define, for each absorber, $\epsilon = |R_{\rm th}-R_{ob}|/|\Delta R|$,
here $R_{\rm th}$  are the CIII/CIV ratios from our models, $R_{\rm ob}$ are the the observed 
values, and $\Delta R$ are the corresponding errors on these measurements (Table~1). 
For each absorber, we  search the entire parameter space of $T$, $n$, $Z$, for a given 
model of ionizing radiation corresponding to the redshift of the absorber. 
For any absorber, a model is deemed  acceptable if $\epsilon \le \sqrt{5}$
(that corresponds to 5$\sigma$-level of the theoretical model).   

Our results for the three models are shown in Tables~2--4. As seen from the tables, 
the models $HM$ and $HM+S$ can fit well for 8 out the 10 data points of 
\citet{agafonova}. However, the $DR$ model is acceptable  for only  six out of 
the ten absorbers. One absorber is fit well by none of the models (absorber 4).

In Tables~2--4 we also show the results for the equilibrium models using
CLOUDY (Figure~3). The results in the two cases have qualitative similarities,
e.g. absorber 4 is ruled out by all models. However, there are also quantitative
differences. Many of the absorbers ruled out by the nonequilibrium models
are now allowed or vice versa.  An interesting departure between the two
results is that the $DR$ model is ruled out for as many as 8 absorbers. 

Next we briefly discuss how our results depend  on the choice of $n$, $T$, and $Z$. 
The preferred range of temperature is generally $\ga \hbox{a few } \times 10^4 \, \rm K$ 
to obtain acceptable solutions for all the absorbers. 
We do not list the best fit values of metallicity in the tables as the metallicities 
in the entire range we considered are allowed. We also note that this is also expected 
if the CIII/CIV ratios were obtained from equilibrium ionization models.
We note that the dependence 
on $n$ is stronger; the value of $n$ is greater than $10^{-3}$cm$^{-3}$ for all the 
acceptable models, except for the DR model. This is in line with the fact that the absorbers 
have densities  comparable to or larger than virialized haloes at $z \simeq 3$, as distinct 
from smaller column density Lyman-$\alpha$ clouds which are only mildly overdense.
In the measurements \citet{agafonova} found a significant HI column densities, 
$N({\rm HI}) \sim 10^{16-17}$cm$^{-2}$. Also in the majority of the absorbers they 
detected singly-ionized carbon and silicon: $N({\rm CII}) \sim 10^{13}$cm$^{-2}$,
$N({\rm SiII}) \sim 10^{12}$cm$^{-2}$, whose ionization potentials are below 13.6~eV. 
But detection of singly-ionized or neutral (like C, O or Si) species depends on the 
filling factor of cold clouds in the ISM of absorber and therefore it is not guaranteed that
such a cold cloud must appear along the line of sight.

Even though this evidence is not conclusive  to rule out the $DR$ model, it is probably an indication that this line of enquiry is likely to better constrain the background far-UV flux.

We note that in absorbing systems with $N({\rm HI}) \ge 10^{16 \hbox{--} 17}$ cm$^{-2}$, the UV background radiation above 54 eV may be attenuated \citep{hm96,mcquinns}. 
That can lead to higher ionic ratio N(CIII)/N(CIV), e.g. for the absorber no. 10 (see Table~1),
thus the absorber no. 10 may suffer from self-shielding. But it is unlikely that the UV flux 
is severely attenuated in the case of other absorbers, because in the extreme case of zero 
flux, the ionic ratio N(CIII)/N(CIV) is expected to be larger than $\sim 20$ for 
$T \le 4 \times 10^4$K, according to Figure~\ref{figc} (the thin dot-dashed line 
for the collisional limit), and all absorbers except no. 10 show a smaller ratio than this, 
suggesting that attenuation of UV flux is unlikely to be a major problem.

\section{Discussions}
Our results suggest that it is possible to explain the ionic ratio without
resorting to additional attenuation of UV flux between 3 and 4 Ryd by
continuum HeII absorption, or a HeII GP effect. \citet{agafonova} used
their data to reconstruct the far-UV background radiation spectrum, and 
claimed a value of $\tau_{GP}({\rm HeII}) \sim 2.5\hbox{--}3$. This was compared
with other direct observations of HeII GP absorption trough, at $z=2.87$,
with opacity $\tau_{GP}({\rm HeII}) \sim 2.09\pm 0.1$ \citep{reimers}.
It is difficult to disentangle the effects of absorption from a continuous
IGM and an ensemble of clouds, as in the case of HI GP effect \citep[see ][]{becker}, 
although \citet{mcquinn} has argued that the case for HeII GP
effect is stronger than this, owing to the low abundance of HeII compared to
HI. In light of the difficulty of identifying true GP effect, our results
of explaining the ionic ratios of CIII/CIV with standard attenuation
from HeII absorption suggest that the observations of \citet{agafonova}
do not necessarily need a HeII GP effect, although it cannot be completely
ruled out.

Our results also clearly show that it is difficult to select one model of far-UV
flux from the observed data. It is possible that the measured ratios are sensitive 
to the local conditions inside the cloud. It is also possible that the far-UV
radiation field is inhomogeneous over the redshift of probe. Available data from 
He~II measurements suggests that the Str\"omgren spheres corresponding to this 
species might just be merging at $z \simeq 3$ \citep{jakobsen,smette}. Numerical 
simulations by \citet{mcquinns} have also shown that the reionization of HeII 
remains patchy even at $z\sim 3$. Therefore, it is entirely conceivable that
it is inhomogeneous reionization and not fluctuation in the density, metallicity 
and temperature that is responsible for the wide range of observed  CIII/CIV ratios. 

A bigger data set involving different ionic ratios
is however needed to address some of these questions in detail,
and confirm the patchiness of the HeII reionization process.
Also, to put better constraints on the spectrum profile we need to study
ionization structure of a cloud coupled with gas dynamics and radiation 
transfer. Radiation transfer effects would particularly affect large
HI column density clouds, and change the predictions of ionic ratios in them,
but these effects are outside the scope of the present paper. Furthermore,
the predicted ionizing background depends on the spectrum and luminosity
function of sources, and this introduces an additional uncertainty
in analysis of the kind we have presented here. It is possible that the effect
of these uncertainties would be mitigated by studying different ionic ratios,
and we will report the advantages of using other ions in a future paper.

\section{Conclusions}
In this paper we used the data of \citet{agafonova} for abundance
ratio of CIII to CIV in quasar absorption systems at $z\sim2.4\hbox{--}2.9$,
to constrain the spectrum of the cosmic UV background radiation. We have
used the spectrum calculated by \citet{mh09} which took into
account the modulation from HeII Lyman series lines, and also models which
assume a large abundance of HeII by artificially increasing the HeII/HI
ratio. Our results can be summarized as follows:

\begin{enumerate}

\item The ratio between CIII/CIV is not a sensitive probe of the attenuation
of the far-UV 
spectrum between 3 and 4 Ryd for standard abundance of HeII in the
IGM. 

\item The observed data do not favor the
models with additional abundance of HeII in the IGM (the "DR", delayed
reionization models).

\item There is a large variation in the fitted values of
density and temperature for the absorption systems. This is indicative
of an inhomogeneous reionization of HeII at these epochs, that has been
suggested from numerical simulations, and our results lend support to these
models.

\end{enumerate}

\acknowledgements
Francesco Haardt is acknowledged for providing the UV background spectrum data and 
necessary explanations. Gary Ferland and CLOUDY community are acknowledged for 
creating of the excellent tool for study of the photo-ionized plasma -- CLOUDY code.
EOV is grateful to Yuri Shchekinov for his help and many useful discussions.
This work is supported by Indo-Russian project (RFBR grant 08-02-91321, DST-India 
grant INT-RFBR-P-10).
EOV is supported by the RFBR (project codes 09-02-00933, 09-02-90726 and 10-02-90705), 
by the Federal Agency of Education (project code RNP 2.1.1/1937) and by the Federal Agency 
of Science and Innovations (project 02.740.11.0247).

\clearpage 

\clearpage 

\clearpage 

\clearpage 

\clearpage 

\clearpage 
\begin{deluxetable}{cccc}
\tablewidth{0pt}
\tablecaption{Carbon ionization ratios for absorption systems}
\tablehead{
\colhead{absorber} & 
\colhead{$z_{abs}$} & 
\colhead{$R=N_{\rm CIII}/N_{\rm CIV}$} & 
\colhead{error in $R$} }
\startdata
1 & 2.379 & 1.438 & 0.160 \\
2 & 2.433 & 6.944 & 0.449 \\
3 & 2.438 & 2.680 & 0.082 \\
4 & 2.568 & 1.000 & 0.083 \\
5 & 2.735 & 3.083 & 0.076 \\
6 & 2.739 & 15.681 & 1.782 \\
7 & 2.741 & 3.158 & 0.499 \\
8 & 2.875 & 4.750 & 0.025 \\
9 & 2.939 & 16.363 & 2.645 \\
10 & 2.944 & 21.765 & 0.311 
\enddata
\end{deluxetable}

\clearpage 
\begin{deluxetable}{ccccccc}
\tablewidth{0pt}
\tablecaption{The best parameters for the model HM. Columns 2--4 
present results for the nonequilibrium model. The next three columns
show the results for the equilibrium model}
\tablehead{
\colhead{absorber} & 
\colhead{T (K)} & 
\colhead{n ($\rm cm^{-3}$)} & 
\colhead{$\epsilon$}  &
\colhead{T (K)} & 
\colhead{n ($\rm cm^{-3}$)} & 
\colhead{$\epsilon$} }
\startdata
1 &  40000 & 0.001 & 4 &  31622 & 0.001 & 2.1 \\
2 & 27200 & 0.01  & 0.1 &  39810 & 0.001 & 8.2\\
3 & 30300 & 0.001 & 0.073 &  31622 & 0.001 & 1.82 \\
4 & 21740 & 0.0001  & 6.9  & 31622 & 0.001  & 9.2 \\
5 & 35460 & 0.001 & 0.02 & 39810 & 0.001 & 2.5  \\
6 & 33300 & 0.01  & 0.027 & 19952 & 0.01 & 0.18 \\
7 & 36370 & 0.001 & 0.005 & 39810 & 0.001 & 0.24  \\
8 & 37920 & 0.01  & 1.36 & 39810 & 0.001 & 58  \\
9 &  12840& 0.01  & 0.006 & 25118 & 0.01 & 0.2  \\
10 & 31730 & 0.01  & 0.15 & 25118 & 0.01 & 1.5
\enddata
\end{deluxetable}

\clearpage 
\begin{deluxetable}{cccccccc}
\tablewidth{0pt}
\tablecaption{Same as Table~2 for the model HM+S}
\tablehead{
\colhead{absorber} & 
\colhead{T (K)} & 
\colhead{n ($\rm cm^{-3}$)} & 
\colhead{$\epsilon$}  &
\colhead{T (K)} & 
\colhead{n ($\rm cm^{-3}$)} & 
\colhead{$\epsilon$} }
\startdata
1 &  39870& 0.001  & 5.7 & 31622 & 0.001 & 2.83\\
2 & 28210 & 0.01  & 0.12 & 39810 & 0.001 & 6.68\\
3 & 34930 & 0.001 & 0.07 & 25118 & 0.001 & 0.85\\
4 & 23800 & 0.0001  & 6.1 & 31622 & 0.001 & 10.7\\
5 & 30200 & 0.001 & 0.1 & 39810 & 0.001 & 2.29\\
6 & 36310 & 0.01  & 0.04 & 19952 & 0.01 & 0.4 \\
7 & 29410 & 0.001 & 0.02 & 39810 & 0.001 & 0.37 \\
8 & 23400 & 0.001  & 0.4 & 39180 & 0.001 & 31\\
9 &  13670& 0.01  & 0.032 & 19552 & 0.01 & 0.012\\
10 & 37070 & 0.01  & 0.015 & 39810 & 0.01 & 0.12
\enddata
\end{deluxetable}

\clearpage 
\begin{deluxetable}{ccccccc}
\tablewidth{0pt}
\tablecaption{Same at Table~2 for the model DR}
\tablehead{
\colhead{absorber} & 
\colhead{T (K)} & 
\colhead{n ($\rm cm^{-3}$)} & 
\colhead{$\epsilon$} &
\colhead{T (K)} & 
\colhead{n ($\rm cm^{-3}$)} & 
\colhead{$\epsilon$} }
\startdata
1 &  39810& 0.0001 & 0.24 & 31622 & 0.001 & 42\\
2 & 31100 & 0.001  & 0.02 & 31622 & 0.001 & 3.1\\
3 & 16840 & 0.0001 & 3.8 & 31622 & 0.001 & 68\\
4 & 39810 & 0.0001  & 5.7 & 31622 & 0.001 &  87 \\
5 & 16840 & 0.0001 & 9.4 & 31622 & 0.001 & 68 \\
6 & 37830 & 0.001  & 0.01 & 39810 & 0.001 & 0.14 \\
7 & 16840 & 0.0001 & 1.6 & 31622  & 0.001 & 10 \\
8 & 39990 & 0.001  & 10.8 & 31622 & 0.001 & 142 \\
9 &  16710& 0.01  & 0.001 & 39810 & 0.001 & 0.16\\
10 & 39630 & 0.01  & 0.015 & 39810 & 0.001 & 18
\enddata
\end{deluxetable}

\end{document}